\documentclass[12pt]{article}

\usepackage{latexsym}
\usepackage{amssymb,amsfonts,amsmath}
\usepackage{graphicx} 
\usepackage{indentfirst}
\usepackage{bbm}
\usepackage{amssymb}
\usepackage{verbatim}
\usepackage{amsmath, amsthm,amssymb}
\usepackage{mathrsfs}
\usepackage{hyperref}
\usepackage{amsfonts}
\usepackage{dsfont}
\usepackage{cite}
\usepackage{xcolor}
\usepackage[multiple]{footmisc}

\parskip=\medskipamount

\newcommand {\cD}{{\cal D}}
\newcommand {\cE}{{\cal E}}

\newcommand {\cL}{{\cal L}}

\newcommand {\cN}{{\cal N}}

\def\a{\alpha}
\def \bi{\bibitem}

\def\b{\beta}

\def\d{\delta}

\def\h{\eta}
\def\f{\phi}
\def\g{\gamma}
\def\G{\Gamma}

\def\j{\psi}
\def\k{\kappa}
\def\l{\lambda}

\def\o{\omega}

\def\q{\theta}

\def\s{\sigma}

\def\x{\xi}
\def\z{\zeta}

\def\J{\Psi}
\def\L{\Lambda}
\def\O{\Omega}

\def\U{\Upsilon}

\def\rd{{\rm d}}
\def\ri{{\rm i}}
\def\re{{\rm e}}

\newcommand{\ad}{{\dot{\alpha}}}                           
\newcommand{\bd}{{\dot{\beta}}}                            
\newcommand{\cDB}{{\bar\cD}}                            

\newcommand{\pa}{\partial}                           
\newcommand{\hf}{\frac12}

\newcommand{\vf}{\varphi}

\newcommand{\be}{\begin{equation}}
\newcommand{\ee}{\end{equation}}
\newcommand{\bea}{\begin{eqnarray}}
\newcommand{\eea}{\end{eqnarray}}
\newcommand{\non}{\nonumber}

\newcommand{\dsR}{{\mathbb R}}

\newcommand{\bm}[1]{\mbox{\boldmath$#1$}}

\def\double #1{#1{\hbox{\kern-2pt $#1$}}}

\newcommand{\mfF}{{\mathfrak F}}
\newcommand{\mfG}{{\mathfrak G}}
\newcommand{\mfH}{{\mathfrak H}}
\newif\ifdtup

\newcommand{\bsubeq}{\begin{subequations}}
\newcommand{\esubeq}{\end{subequations}}

\numberwithin{equation}{section}

\newcommand{\sSU}{\mathsf{SU}}

\newcommand{\sU}{\mathsf{U}}

\begin{document}

\begin{titlepage}
\begin{flushright}
August, 2023 \\
Revised version: November, 2023
\end{flushright}
\vspace{5mm}

\begin{center}
{\Large \bf 
New duality-invariant models for nonlinear supersymmetric electrodynamics} 
\end{center}

\begin{center}

{\bf Sergei M. Kuzenko and Jake C. Stirling} \\
\vspace{5mm}

\footnotesize{
{\it Department of Physics M013, The University of Western Australia\\
35 Stirling Highway, Perth W.A. 6009, Australia}}  
~\\
\vspace{2mm}
~\\
Email: \texttt{ 
sergei.kuzenko@uwa.edu.au, jake.stirling@research.uwa.edu.au}\\
\vspace{2mm}

\end{center}

\begin{abstract}
\baselineskip=14pt
We propose a new family of $\mathsf{U}(1)$ duality-invariant models for nonlinear ${\cal N}=1$ supersymmetric electrodynamics coupled to supergravity. It includes the Cribiori-Farakos-Tournoy-van Proeyen supergravity-matter theory for spontaneously broken local supersymmetry with a novel Fayet-Iliopoulos term without gauged $R$-symmetry. We present superconformal duality-invariant models, as well as new $\mathsf{U}(1)$ duality-invariant models for spontaneously broken local supersymmetry.
\end{abstract}
\vspace{5mm}

\vfill

\vfill
\end{titlepage}

\newpage
\renewcommand{\thefootnote}{\arabic{footnote}}
\setcounter{footnote}{0}


\allowdisplaybreaks


\section{Introduction} \label{Section1}

The general theory of duality invariance for nonlinear models with $\cN=1$ and $\cN=2$ vector supermultiplets was developed in 2000 \cite{KT1,KT2} and soon after  extended to the locally supersymmetric case  \cite{KMcC,KMcC2,K12}. This formalism
is a generalisation of the classic results on the structure of self-dual
models for nonlinear electrodynamics in four dimensions
\cite{GZ1,GR1,GR2,GZ2,GZ3} (see \cite{KT2,AFZ} for a review) 
 in conjunction with the self-duality properties
\cite{BMZ,ABMZ} of the $\cN=1$ supersymmetric Born-Infeld action \cite{CF,BG,RT} and its generalisations.

At the turn of the millennium, the main motivation to study supersymmetric self-dual systems was the existence of deep yet (still) mysterious connections between nonlinear self-duality and supersymmetry. These are:
\begin{itemize}
\item
In the case of partial spontaneous
supersymmetry breakdown $\cN=2 \to \cN=1$, the Maxwell-Goldstone multiplet \cite{BG,RT}
(coinciding with the $\cN=1$ supersymmetric Born-Infeld
action \cite{CF}) and the tensor Goldstone multiplet
\cite{BG2,RT} were shown  in \cite{KT1,KT2} to be invariant under $\sU(1)$ duality rotations.\footnote{The  Maxwell-Goldstone multiple for partial $\cN=2 \to \cN=1$ supersymmetry breaking has also been extended \cite{KT-M16} to the following maximally supersymmetric backgrounds: (i)  ${\mathbb R} \times S^3$; (ii) ${\rm AdS}_3 \times {\mathbb R}$; and (iii) a supersymmetric plane wave. This theory possesses $\sU(1)$ duality invariance.} 

 \item
 Extending the earlier incomplete proposal of \cite{Ketov}, 
 it was suggested in \cite{KT2} that the Maxwell-Goldstone multiplet for 
 partial  $\cN=4 \to \cN=2$ supersymmetry breakdown (proposed to be the $\cN=2$ supersymmetric Born-Infeld action) is a unique $\cN=2$ vector multiplet theory with the following properties: (i) it possesses $\sU(1)$ duality invariance; and (ii) it is invariant under a nonlinearly realised central charge bosonic symmetry. Within the
perturbative approach to constructing the $\cN=2$ supersymmetric Born-Infeld action 
elaborated in  \cite{KT2}, the uniqueness of the action was demonstrated to order $W^{10}$ in powers of the chiral superfield strength $W$.
A year later, a powerful formalism of nonlinear realisations for
the partial  $\cN=4 \to \cN=2$ supersymmetry breaking was developed \cite{BIK1}
which supported the uniqueness of the $\cN=2$ supersymmetric Born-Infeld action 
and reproduced \cite{BIK2} the perturbative results of \cite{KT2}.
Further progress towards the construction of the $\cN=2$ supersymmetric Born-Infeld action has been achieved in \cite{BCFKR,CK,IZ4}. 

\item 
The $\cN=4$ super Yang-Mills (SYM) theory is believed to be self-dual 
\cite{MO,Osborn} (see also \cite{Sen}). This conjecture was  put forward in the late 1970s as a duality between the conventional and soliton sectors  of the theory. 
In 1998 it was suggested \cite{GKPR}
that self-duality might be realised in terms of a low-energy 
effective action of the theory on its Coulomb branch.\footnote{This was inspired in part by the Seiberg-Witten theory \cite{SW1,SW2} and also by the AdS/CFT correspondence 
\cite{Maldacena}.}
Here the gauge group $\sSU(N)$ is spontaneously 
broken to $\sSU(N-1) \times \sU(1)$ and the dynamics 
is described by a nonlinear $\cN=2$ {\it superconformal} action for the  $\cN=2$ vector multiplet associated with the $\sU(1)$ factor 
of the unbroken group. 
Two different realisations of self-duality for the $\cN=4$ SYM effective action
in $\cN=2$ superspace were  proposed:
(i) self-duality under Legendre transformation
\cite{GKPR}; and (ii) self-duality under 
$\sU(1)$ duality rotations \cite{KT1}.\footnote{It was also conjectured by Schwarz 
\cite{Schwarz:2013wra} that the world-volume action of a probe D3-brane in an 
$\rm AdS_5 \times S^5$ background of type IIB superstring theory, with one unit of flux, can be reinterpreted as the exact (or highly) effective action for 
$\sU(2)$ $\cN = 4$ super Yang-Mills theory on the Coulomb branch.
}
Both proposals have not yet been derived from first principles, although each of them is consistent with the one-loop \cite{PvU, G-RR, BK98, BBK,LvU} and
two-loop \cite{K2004} calculations.

\item For a large class of {\it nonlinear} $\sU(1)$ duality-invariant models for $\cN=1$ supersymmetric electrodynamics \cite{KT1}, it was demonstrated \cite{KMcC2} that the component fermionic action, which is obtained by switching the bosonic fields off, is equivalent (modulo a nonlinear field redefinition) to the Akulov-Volkov action for the Goldstino \cite{VA,AV}.\footnote{An explanation of this result was given in \cite{K-FI}.} 
\end{itemize}
A new motivation to revisit the general structure of $\sU(1)$ duality-invariant models 
for the $\cN=1$ vector multiplet emerged five years ago when Cribiori et al.
\cite{Cribiori:2017laj} discovered a novel Fayet-Iliopoulos term in supergravity without gauged $R$-symmetry. In order to explain this motivation, it is pertinent to give a summary of the formalism introduced in \cite{KT1} and its generalisation advocated in \cite{ILZ}.

Let  $S[W , \bar W]$ be the action describing the dynamics of a single vector supermultiplet 
in Minkowski superspace. It is assumed that the action is a functional of the gauge-invariant 
chiral spinor field strength $W_\a = -\frac 14 \bar D^2 D_\a V$ and its conjugate
$\bar W_\ad = -\frac 14  D^2 \bar D_\ad V$, where $V=\bar V$ is a gauge prepotential
\cite{FZ}. 
In order for this theory to possess $\sU(1)$ duality invariance, 
 the action must be a solution of the so-called self-duality equation \cite{KT1}
\bea
{\rm Im} \int \rd^4 x \rd^2 \q  \, \Big\{ W^\a W_\a  +M^\a M_\a \Big\} =0~,
\qquad M_\a := -2\ri  \,\frac{\d }{\d W^\a}\,S[W , {\bar W}]~.
\label{1.1}
\eea

Since the equation \eqref{1.1} is nonlinear, its solutions are difficult to generate.  
Inspired by the bosonic approach due to Ivanov and Zupnik
 \cite{IZ_N3,IZ1,IZ2,IZ3},  new formulations were developed for $\cN=1$ and 
$\cN=2$ supersymmetric duality-invariant theories coupled to supergravity ten years ago \cite{K13}. The method makes use of an auxiliary unconstrained chiral superfield (a spinor in the $\cN=1$ case and a scalar for $\cN=2$) and is characterised by the  fundamental property that $\sU(1)$ duality invariance is equivalent to the manifest $\sU(1)$ invariance of the self-interaction. In the $\cN=1$ rigid supersymmetric case,
analogous results were independently obtained in \cite{ILZ}.

It is assumed in \eqref{1.1} that $W_\a$ is an unrestricted chiral spinor superfield, and the action $S[W , \bar W]$ is ``analytically'' continued from the original functional, which  depends on $W_\a$ satisfying the Bianchi identity $D^\a W_\a = \bar D_\ad \bar W^\ad$, to a functional of the unrestricted chiral spinor $W_\a$. Such a continuation is obviously not unique, and additional conditions are required to fix it. For instance,  
consider a supersymmetric theory of the form
\bea
S[W,{\bar W}] &=&
\frac{1}{4} \int  \rd^4 x \rd^2 \q   \, W^2 +{\rm c.c.} + \frac{1}{4}  \int \rd^4 x \rd^2 \q \rd^2\bar \q \,
W^2\,\bar W^2
\O\left(u,\bar u, D^\a W_\a \right)~,
\label{1.2}
\eea
where $W^2 =W^\a W_\a$ and $u = \frac{1}{8} D^2  W^2$. A possible way to extend this functional to the case when the Bianchi identity $D^\a W_\a = \bar D_\ad \bar W^\ad$ is no longer required, consists in replacing $\O\left(u,\bar u, D^\a W_\a \right)  \to 
\O\left(u,\bar u, \g D^\a W_\a +\bar \g \bar D_\ad \bar W^\ad \right)$, for a complex parameter $\g$ such that $\g +\bar \g=1$. 

The ambiguity with analytic continuation is naturally resolved for the family of nonlinear models studied in \cite{KT1}:
\bea
S[W,{\bar W}] &=&
\frac{1}{4} \int  \rd^4 x \rd^2 \q   \, W^2 +{\rm c.c.} + \frac{1}{4}  \int \rd^4 x \rd^2 \q \rd^2\bar \q \,
W^2\,\bar W^2
\L\left(u,\bar u\right)~.
\label{SDSED}
\eea
Here the self-duality equation \eqref{1.1} implies that 
\bea
{\rm Im}\Big\{\G-\bar u\G^2 \Big\}=0~,\qquad \G:=\pa_u(u\L)~.
\label{1.3}
\eea
This equation coincides in form with that arising in $\sU(1)$ duality-invariant nonlinear electrodynamics $L(F_{ab})$ (see \cite{KT2} for the technical details),
\bea
{\rm Im}  \left\{ \frac{\pa (\o \, \L) }{\pa \o}
- \bar{\o}\,
\left( \frac{\pa (\o \, \L )  }{\pa \o} \right)^2 \right\} = 0~,
\label{GZ4}
\eea
provided the Lagrangian is expressed in terms of invariants of the electromagnetic field 
\begin{subequations}\label{SDED}
\bea
L (F_{ab}) = -\hf \, \Big( \o + \bar{\o} \Big) +\o \, \bar{\o} \; \L (\o, \bar{\o} )~,
\eea
where we have introduced
\bea
\o = \a + {\rm i} \, \b~, \qquad \a = \frac{1}{4} \, F^{ab} F_{ab}~, \qquad \quad
\b = \frac{1}{4} \, F^{ab} \tilde{F}_{ab} ~.
\label{omega}
\eea
\end{subequations}
Therefore, every $\sU(1)$ duality-invariant model for nonlinear electrodynamics \eqref{SDED}
 possesses  the $\cN=1$ supersymmetric extension \eqref{SDSED}
which is also $\sU(1)$ duality invariant. 

Instead of worrying about a procedure for analytic continuation to start with, one can follow a different path proposed by Ivanov, Lechtenfeld and Zupnik \cite{ILZ}. Their starting point is the assumption that some procedure of analytic continuation has been chosen, and for the unconstrained chiral spinor $W_\a$ the action reads  
\bea
S[W,{\bar W}] &=&
\frac{1}{4} \int  \rd^4 x \rd^2 \q  \, W^2 +{\rm c.c.} + \frac{1}{4}  \int \rd^4 x \rd^2 \q \rd^2\bar \q \,
W^2\,\bar W^2
\L\left(u,\bar u,p,\bar p\right)~,
\eea
where $ p:=D^\a W_\a$. For such a model to be $\sU(1)$ duality-invariant, the action must satisfy the equation \eqref{1.1}, and the latter implies that 
\bea
{\rm Im}\Big\{\G-\bar u\G^2+2u\bar u(\pa_p\L)^2\Big\}=0~.
\label{SDE2.8}
\eea
To the best of our knowledge, no solution of the self-duality equation \eqref{SDE2.8} has so far been found with $\pa_p \L \neq 0$.
In the present paper, we  propose a family of such solutions.
 
Functionals of the type \eqref{1.2} naturally appear as
low-energy effective actions in quantum supersymmetric gauge theories, see, e.g., \cite{MG,PB,BKT}. However, the combination $ D^\a  W_\a$
is nothing but the free equation of motion of the $\cN=1$
vector multiplet. It is known that those contributions to the effective action, which contain
factors of the classical equations of motion, are ambiguous. That is why only action functionals of the form 
\eqref{SDSED} were studied in Refs. \cite{KT1,KT2}. On the other hand, there may exist microscopic models that involve  $ D^\a  W_\a$ in the superfield Lagrangian, and then one is forced to deal with models \eqref{1.2}.
This is exactly the case with the model of \cite{Cribiori:2017laj} proposed to describe a novel Fayet-Iliopoulos term in supergravity without gauged $R$-symmetry. Restricted to a flat superspace background, the corresponding vector multiplet action is 
\bea
S[W,{\bar W}] &=&
\frac{1}{4} \int  \rd^4 x \rd^2 \q  \, W^2 +{\rm c.c.} + \frac{\z}{4}  \int \rd^4 x \rd^2 \q \rd^2\bar \q \,
\frac{W^2\bar W^2} {u\bar u}D^\a W_\a~,
\label{CFTVP}
\eea
with $\z$ a coupling constant. For this action to be well defined, $u = \frac{1}{8} D^2  W^2$ should be nowhere vanishing. This requirement is consistent with the equations of motion, since the auxiliary field of the vector multiplet develops a non-zero VEV on-shell. 

The model \eqref{CFTVP} was demonstrated in \cite{Cribiori:2017laj} to be self-dual under a superfield Legendre transformation. In this paper, we will prove that \eqref{CFTVP} possesses $\sU(1)$ duality invariance. 
It is a general property of $\sU(1)$ duality-invariant theories that every solution of the self-duality equation 
\eqref{1.1} is self-dual under the Legendre transformation \cite{KT1,KT2}. Therefore, our analysis in this paper implies the self-duality property established in \cite{Cribiori:2017laj}.

Before turning to the technical part of this paper, it is necessary to point out several recent developments. Although the idea to combine $\sU(1)$ duality invariance with 
$\cN=2$ superconformal symmetry was put forward in 2000 \cite{KT1}, the first duality-invariant and (super)conformal theories  have only recently been derived in closed form for $\cN<2$. Bandos, Lechner, Sorokin and Townsend \cite{BLST} 
constructed the so-called ModMax theory, which is a unique nonlinear duality-invariant and conformal extension of Maxwell's equations (see \cite{Kosyakov} for a related analysis).
Its $\cN=1$ supersymmetric extension was given in \cite{BLST2,K21}. 
Ref. \cite{K21} also derived  the $\cN=2$ superconformal $\sU(1)$ duality-invariant model proposed to describe the low-energy effective action for $\cN=4$ super-Yang-Mills theory, thus completing the program initiated in \cite{KT1}. 
Duality-invariant (super)conformal higher-spin models were constructed for $\cN \leq 2$ in \cite{KR21-2} on arbitrary conformally flat backgrounds. 
A supersymmetric nonlinear $\sigma$-model analogue of the ModMax theory
was constructed in \cite{KMcA23} building on the concept of self-dual supersymmetric nonlinear $\sigma$-models \cite{KMcA}.
Supersymmetric duality-invariant theories have found numerous applications in the framework of $T \bar T$ deformations, see \cite{FT-M1,FT-M2,FT-M3} and references therein. A remarkable relation has been established between helicity conservation for the tree-level scattering amplitudes and the electric-magnetic duality\cite{Novotny}. 

This paper is organised as follows. In section \ref{Section2} we briefly review the $\cN=1$ results of \cite{KT1,KMcC,ILZ} and then present our new family of $\mathsf{U}(1)$ duality-invariant models for nonlinear ${\cal N}=1$ supersymmetric electrodynamics coupled to supergravity. In section \ref{Section3} we provide a brief review of the $\cN=1$ auxiliary superfield formulation of \cite{K13} and then recast the novel features of \cite{ILZ} (as compared with \cite{K13}) in a locally supersymmetric framework. Section \ref{Section4} is devoted to superconformal $\sU(1)$ duality-invariant models. Section \ref{section5} discusses the obtained results and introduces new $\mathsf{U}(1)$ duality-invariant models for spontaneously broken local supersymmetry.

Our two-component spinor notation and conventions follow 
\cite{Ideas}, and are similar to those adopted in \cite{WB}.  We make use of the 
Grimm-Wess-Zumino superspace geometry \cite{GWZ} as described in \cite{Ideas,KRT-M}. 


\section{Duality-invariant supersymmetric models} \label{Section2}

We consider a dynamical system describing an Abelian $\cN = 1$
vector multiplet in curved superspace and denote by $S[W , {\bar W}]$ 
the corresponding action functional. The action is assumed to depend
on the  chiral spinor field strength $W_\a$
and its conjugate ${\bar W}_\ad$ which are constructed 
in terms of a real unconstrained gauge prepotential $V$ \cite{WZ} as
\bea
W_\a = -\frac{1}{4}\,  (\bar \cD^2 -4R) 
\cD_\a  V~, \qquad \bar \cD_\bd W_\a=0~.
\eea 
Here $\cD_\a$ and $\bar \cD_\ad$ are the spinor covariant derivatives in curved superspace, and $R$ is the chiral scalar torsion tensor.\footnote{Our normalisation of the torsion tensors of the Grimm-Wess-Zumino geometry follows \cite{Ideas,KRT-M}.} 
The prepotential is defined modulo gauge transformations
\bea
\d_\x V = \x + \bar \x ~, \qquad \bar \cD_\ad \x =0~, 
\eea
such that $\d_\x W_\a =0$.
The gauge-invariant  field strengths $W_\a$ and ${\bar W}_\ad$  obey
the Bianchi identity
\bea
\cD^\a W_\a = \bar \cD_\ad {\bar W}^\ad~,
\label{eq:bianchi}
\eea
and thus $W_\a$ is a reduced chiral superfield. 
We assume that $S[W , {\bar W}]$ does not involve the combination $\cD^\a W_\a $ as an independent variable, and therefore it
can unambiguously be defined
as a functional of a {\it general} 
chiral superfield $W_\a$ and its conjugate ${\bar W}_\ad$.
Then, introducing a covariantly chiral spinor superfield $M_\a$, 
\bea
{\rm i}\,M_\a := 2\, \frac{\d }{\d W^\a}\,S[W , {\bar W}]~,  \qquad \bar \cD_\bd M_\a=0~,
\label{2.3M}
\eea
the equation of motion for $V $ is 
\bea
\cD^\a M_\a = \cDB_\ad {\bar M}^\ad~.
\label{eq:eom}
\eea
The variational derivative $\d S/\d W^\a $ in \eqref{2.3M} is defined by 
\bea
\d S =  \int \rd^4 x \,{\rm d}^2 \q \,\cE\, \d W^\a \frac{\d S}{\d W^\a}~+~{\rm c.c.}~,
\eea
where $\cE$ denotes the chiral integration measure, and $W^\a$ is
assumed to be an unrestricted covariantly chiral spinor.

Since the Bianchi identity (\ref{eq:bianchi}) and the equation of
motion (\ref{eq:eom}) have the same functional form, one may
consider $\sU(1)$ duality rotations
\bea
\d W_\a = \l M_\a ~, 
\qquad \d M_\a = - \l W_\a~,
\label{DualRot}
\eea
with $\l \in {\mathbb R}$ a constant parameter. The condition for duality invariance is the so-called self-duality equation 
\bea
{\rm Im} \int \rd^4 x \rd^2 \q  \,\cE \Big\{ W^\a W_\a  +M^\a M_\a \Big\} =0~,
\label{SDE1}
\eea
in which $W_\a$ is chosen to be a general chiral spinor superfield.

In what follows, we shall treat $W_\a$ and $\bar W_\ad$ as unconstrained chiral superfields which are not subjected to the Bianchi identity (\ref{eq:bianchi}). 
Now let us introduce the following general model for nonlinear $\cN=1$ supersymmetric electrodynamics 
\bea
S[W,{\bar W}] &=&
\frac{1}{4} \int  \rd^4 x \rd^2 \q  \,\cE \, W^2 +{\rm c.c.} + \frac{1}{4}  \int \rd^4 x \rd^2 \q \rd^2\bar \q \,E \,
W^2\,\bar W^2\,
\L\left(u,\bar u,p,\bar p\right)~,
\label{multipletaction}
\eea
where the complex variables $u$ and $p$ are defined by
\bea
u  := \frac{1}{8} (\cD^2 - 4  \bar R)  W^2~, \qquad p:=\cD^\a W_\a~.
\eea
For this model the self-duality equation (\ref{SDE1}) amounts to
\bea\label{SDE2}
{\rm Im} \int \rd^4 x \rd^2 \q \rd^2\bar \q \,E \,
W^2\,\bar W^2\Big\{\G-\bar u\G^2+2u\bar u(\pa_p\L)^2\Big\}=0~,\qquad \G:=\pa_u(u\L)~.
\eea
In this equation the covariantly chiral spinor $W_\a$ has to be completely arbitrary, 
and therefore we conclude that it suffices for the equation \eqref{SDE2.8} to hold.

A super-Weyl invariant equivalent of the model (\ref{multipletaction}) is given by 
\bea
\label{superweylmultipletaction}
S[W,{\bar W};\U] &=&
\frac{1}{4} \int  \rd^4 x \rd^2 \q  \,\cE \, W^2 +{\rm c.c.}\non\\
&& + \frac{1}{4}  \int \rd^4 x \rd^2 \q \rd^2\bar \q \,E \,
\frac{W^2\,\bar W^2}{\U^2}\L\left(\frac{u}{\U^2},\frac{\bar u}{\U^2},\frac{p}{\U},\frac{\bar p}{\U}\right)~,
\eea
where $\U$ is a nowhere vanishing real scalar with the super-Weyl transformation
\bea\label{compensatortrans}
\d_\s\U=(\s+\bar\s)\U~,\qquad \bar\cD_\bd\s=0~,
\eea
with $\s$ being the super-Weyl parameter.\footnote{See, e.g., \cite{KRT-M} for a review of super-Weyl transformations within the Grimm-Wess-Zumino geometry.}
The transformation law (\ref{compensatortrans}) implies that (\ref{superweylmultipletaction}) is super-Weyl invariant. 
One may readily check that if $\L\left(u,\bar u,p,\bar p\right)$ is a solution of the equation 
\eqref{SDE2.8}, then 
\bea
\widetilde{\L}\left(u,\bar u,p,\bar p ; \U\right)
:= \frac{1}{\U^2}\L\left(\frac{u}{\U^2},\frac{\bar u}{\U^2},\frac{p}{\U},\frac{\bar p}{\U}\right)
\eea
is also a solution of the same equation, 
and thus the model \eqref{superweylmultipletaction}
is $\sU(1)$ duality-invariant. 

The family of $\sU(1)$ duality-invariant theories analysed in \cite{KMcC2,K13} is given by 
\bea
\label{superweylmultipletaction-old}
S[W,{\bar W};\U] &=&
\frac{1}{4} \int  \rd^4 x \rd^2 \q  \,\cE \, W^2 +{\rm c.c.}
\non\\
&& 
+ \frac{1}{4}  \int \rd^4 x \rd^2 \q \rd^2\bar \q \,E \,
\frac{W^2\,\bar W^2}{\U^2}\L\left(\frac{u}{\U^2},\frac{\bar u}{\U^2}\right)~.
\eea
In this case the self-duality equation \eqref{SDE2.8} turns into \eqref{1.3}.

We now present several simple solutions to the self-duality equation \eqref{SDE2.8}.
Consider a polynomial interaction homogeneous in $p$ and $\bar p$,
\bea\label{polynomialint}
\L^{(n)}(u,\bar u,p,\bar p)=\frac{1}{u\bar u}\sum_{k=0}^{n}a_kp^k\bar{p}^{n-k}~,\qquad a_{n-k}=\bar{a}_k=a_k~.
\eea
The self-duality equation (\ref{SDE2.8}) amounts to the following conditions on the coefficients $a_k$:
\begin{subequations}\label{polynomialcond}
\bea
k^2a_k^2-(n-k+1)^2 a_{n-k+1}^2&=&0~,\qquad k=1,\ldots,n~, \\
kla_ka_l-(n-k+1)(n-l+1)a_{n-k+1}a_{n-l+1}&=&0~,\qquad l\neq k~.
\eea
\end{subequations}
We end up with 
\bea\label{coefficientrule}
a_j=\frac{n!}{j!(n-j)!}a_n~,\qquad j=\left\{
\begin{aligned}
& 1,\ldots,\frac{n}{2}\quad \text{for $n$ even}\\
&1,\ldots,\frac{n+1}{2}\quad \text{for $n$ odd}\\
\end{aligned}\right.~.
\eea

As a result, the interaction (\ref{polynomialint}) can be written in the form
\begin{subequations}\label{self-coupling}
\bea
\L^{(n)}(u,\bar u,p,\bar p)=\frac{\z}{u\bar u}\left( \frac{p+\bar p}{2}\right)^n~, 
\label{2.19a}
\eea
with $\z$ a coupling constant. This solution of the self-duality equation (\ref{SDE2.8}) has an obvious polynomial generalisation
\bea
\L^{[n]}(u,\bar u,p,\bar p)=  \frac{1}{u\bar u} \sum_{k=0}^n \z_k \left( \frac{p+\bar p}{2}\right)^k~. 
\label{2.19b}
\eea
\end{subequations}
In order for the couplings \eqref{self-coupling} to be well-defined, the descendant $u$ should be nowhere vanishing. In order for this condition to be consistent with the equations of motion, 
in general we should require the coefficient $\z_1$ to be non-zero. We will come back to a discussion of this issue in section \ref{section5}.\footnote{At this stage we are only interested  in generating solutions of the self-duality equation (\ref{SDE2.8}) without worrying about consistency issues.}

Of special interest is the $n=1$ case in \eqref{2.19a}, since it corresponds to the model introduced in \cite{Cribiori:2017laj} and describes 
a Fayet-Iliopoulos term in supergravity without gauged $R$-symmetry, eq. \eqref{CFTVP}.\footnote{
Models with more general  Fayet-Iliopoulos terms described in \cite{K18} do not appear to be duality invariant.}

The duality-invariant model constructed, eq. \eqref{self-coupling}, can be generalised as follows 
\bea
\L(u,\bar u,p,\bar p)=\frac{1}{u\bar u} {\mathfrak D}\left( \frac{p+\bar p}{2}\right)~, 
\label{self-coupling2}
\eea
where $\mathfrak D(x) $ is a real function of a real argument. It is easy to see that 
\eqref{self-coupling2} is a solution to the self-duality equation \eqref{SDE2.8}.
This is a special case of more general solutions 
\bea
\L(u,\bar u,p,\bar p)=\L (u, \bar u) 
+ \frac{1}{u\bar u} {\mathfrak D}\left( \frac{p+\bar p}{2}\right)~,
\eea
where $\L(u, \bar u)$ is an arbitrary solution of the self-duality equation \eqref{1.3}.

More general solutions of the self-duality equation \eqref{SDE2.8} may be obtained. Let $\L(u,\bar u; \g)$ be a solution of \eqref{1.3}, which depends on a real duality-invariant parameter $\g$. Then the following self-interaction
\bea
\L\big(u,\bar u,p,\bar p\big):= \L\left(u,\bar u; {\mathfrak D}\Big( \frac{p+\bar p}{2}\Big) \right)
\label{2.21}
\eea
is a solution of the self-duality equation \eqref{SDE2.8}, for any real function 
$\mathfrak D(x) $ of a real variable. A converse statement also holds. 
If  $\L\big(u,\bar u,p,\bar p\big):={\bm \L}(u,\bar u; \hf p +\hf \bar p)$
is a solution of  \eqref{SDE2.8}, then $\L\big(u,\bar u;\g\big):= {\bm \L}(u,\bar u, \g)$ is a solution of the 
self-duality equation \eqref{1.3}, with $\g$ being the duality-invariant parameter. 

It should be emphasised that the parameter $\g$ in $\L(u,\bar u; \g)$ above is duality invariant. However, 
the combination $(p+\bar p)$ in \eqref{2.21} is not duality invariant, even on-shell in the nonlinear case, as follows from the infinitesimal duality transformation \eqref{DualRot}. The duality transformation acts on all components of the vector multiplet, and not only on the Maxwell tensor. 

As an example, let us consider the supersymmetric ModMax theory  \cite{BLST2,K21}
\bea
S_{\rm sModMax} &=&
\frac{1}{4}  \int  \rd^4 x \rd^2 \q  \,\cE \,  W^2 \cosh \g +{\rm c.c.}
+ \frac{1}{4} \int \rd^4 x \rd^2 \q \rd^2\bar \q \,E \,
\frac{W^2\,{\bar W}^2}{\sqrt{u\bar u} } \sinh \g  
~,~~
\label{ModMax}
\eea
where $\g $ is the duality-invariant parameter. With this form of the action, the deformation prescription \eqref{2.21}, 
that is
$\g \to {\mathfrak D}\Big( \frac{p+\bar p}{2}\Big) $ is not applicable. However, \eqref{ModMax} can be rewritten in the  alternative form 
\begin{subequations}
\bea
S_{\rm sModMax} &=&
\frac{1}{4} \int  \rd^4 x \rd^2 \q   \, \cE W^2 +{\rm c.c.} + \frac{1}{4}  \int \rd^4 x \rd^2 \q \rd^2\bar \q \,E
W^2\,\bar W^2
\L_{\rm SC} \left(u,\bar u ; \g \right)~,~~
\label{ModMax2}
\eea
where we have introduced 
\bea
\L_{\rm SC}(u, \bar u; \g) = 
\hf (1-\cosh \g ) \Big( \frac{1}{u} 
+\frac{1}{\bar u} \Big)+ \frac{\sinh \g}{\sqrt{u\bar u} } ~.
\eea
\end{subequations}
Since $\L_{\rm SC}(u, \bar u; \g) $ is a solution of \eqref{1.3},  the deformation prescription \eqref{2.21} is applicable and leads to a solution of the self-duality equation \eqref{SDE2.8}.


\section{Auxiliary superfield formalism} \label{Section3}

In this section we describe an alternative formulation for the self-dual models of the $\cN=1$ vector multiplet described in the previous section. In particular, we extend the results obtained in \cite{ILZ} to the case of curved superspace.  

We start with a brief summary of the construction given in \cite{K13}.
Consider an auxiliary action of the form
\bea
S[W,\bar W, \eta, \bar \eta]= \int \rd^4 x \rd^2 \q  \,\cE \Big\{ \eta W -\hf \eta^2 - \frac{1}{4} W^2\Big\} 
+{\rm c.c.}  + {\mathfrak S}_{\rm int} [\eta, \bar \eta]~.
\label{3.1}
\eea
Here the spinor superfield $\eta_\a$ 
is constrained to be covariantly chiral, $\bar \cD_\bd \eta_\a =0$, 
but otherwise it is completely arbitrary. 
By definition, the second term on the right, $ {\mathfrak S}_{\rm int} [\eta, \bar \eta]$,
contains cubic, quartic and higher powers of $\eta_\a$ and its conjugate. 

The above model is equivalent to a theory with action 
\bea
S[W,\bar W] =    \frac{1}{4}\int \rd^4 x \rd^2 \q  \,\cE  \, W^2 
+{\rm c.c.}  + S_{\rm int} [W, \bar W]~,
\label{VMaction}
\eea
describing the dynamics
of the vector multiplet. Indeed, under reasonable assumptions the equation of motion for $\eta^\a$ 
\bea
W_\a = \eta_\a - \frac{\d}{\d \eta^\a} {\mathfrak S}_{\rm int} [\eta, \bar \eta]
\eea
allows one to express $\eta_\a$ as a functional 
of $W_\a$ and its conjugate, $\eta_\a = \J_\a[W, \bar W]$. Plugging this functional and its conjugate into  \eqref{3.1} leads to  a vector-multiplet model 
of the form \eqref{VMaction}. If $S[W,\bar W]$ is a solution of the self-duality equation \eqref{SDE1}, then the self-interaction $ {\mathfrak S}_{\rm int} [\eta, \bar \eta]$ in \eqref{3.1} proves to be invariant 
under rigid $\sU(1)$ phase transformations of $\eta_\a$ and its conjugate,
\bea
{\mathfrak S}_{\rm int} [ \re^{-\ri \l}  \eta, \re^{\ri \l} \bar \eta] = {\mathfrak S}_{\rm int} [\eta, \bar \eta] ~, 
\qquad \l \in {\mathbb R}~.
\label{3.4}
\eea
The duality rotation \eqref{DualRot} acts on the chiral spinor $\eta_\a$ as
\bea
\d \eta_\a = - \ri \l \eta_\a~,
\label{DualTransformationL}
\eea
see \cite{K13} for the technical details.

We now restrict our attention to a subclass of the models \eqref{3.1} of the form:
\begin{subequations}\label{auxiliaryform}
	\bea\label{auxiliaryaction}
	S[W,\bar W, \eta, \bar \eta]&=& \int{\rm d}^4 x \rd^2\q\,\cE \Big\{ \eta W -\hf \eta^2 - \frac{1}{4} W^2\Big\} 
	+{\rm c.c.} \non \\
	&& + \frac{1}{4} \int \rd^4 x \rd^2 \q \rd^2\bar \q \,E \, \eta^2 \bar \eta^2 
	{\mathfrak F}(v,\bar v,q,\bar q)~,
	\eea
	in which
	\bea
	v:= \frac{1}{8} (\cD^2 -4\bar R) \eta^2~, \qquad q:=\cD^\a\h_\a
	\eea
\end{subequations}
and $\mathfrak{F}(v,\bar v,q,\bar q)$ is a real function. 
Such models in a flat background were analysed in \cite{ILZ}.

We aim to integrate out from (\ref{auxiliaryaction}) the auxiliary spinor variables $\h_\a$ and $\bar \h_\ad$ in order to bring the action to the form (\ref{multipletaction}). The equation of motion for $\h^\a$ is
\bea\label{EoM}
W_\a &=& \eta_\a \left\{ 1 + \frac{1}{8} (\bar \cD^2 -4R) 
\Big[ {\bar \eta}^2 \Big(\mathfrak F  + \frac{1}{8} (\cD^2 -4\bar R) 
\big( \eta^2\, \pa_v \mathfrak F \big) \Big) \Big] \right\} \non \\
&&-\frac{1}{16}(\bar \cD^2-4R)\left[\bar \h^2\cD_\a(\h^2\pa_q\mathfrak{F})\right]~.
\eea
Its immediate implications are
\begin{subequations}\label{auxidentities}
\bea
\h W &=& \h^2\left\{1+\frac{1}{8}(\bar \cD^2-4R)\Big[\bar \h^2\Big(\pa_v(v\mathfrak{F})+\frac{1}{2}q\pa_q\mathfrak{F}\Big)\Big]\right\}~,\\
W^2 &=& \h^2\Bigg\{\Big[1+\frac{1}{8}(\bar \cD^2-4R)\left\{\bar \h^2\pa_v(v\mathfrak{F})\right\}\Big]^2+\frac{1}{8}(\bar \cD^2-4R)\left(\bar\h^2 q\pa_q\mathfrak{F}\right) \label{W^2}\non \\
&&+\frac{1}{64}(\bar \cD^2-4R)\left(\bar\h^2\pa_v(v\mfF)\right)(\bar\cD^2-4R)\left(\bar\h^2 q\pa_q\mfF\right) \non \\
&&+\frac{1}{128}(\bar\cD^2-4R)\left(\bar\h^2(\cD^\a\h^\b)\pa_q\mfF\right)(\bar\cD^2-4R)\left(\bar\h^2(\cD_\a\h_\b)\pa_q\mfF\right)\Bigg\}~,\\
W^2\bar W^2&=&\h^2\bar\h^2\mfH\bar\mfH~\label{W^2Wbar^2},
\eea
\end{subequations}
where we have introduced 
\bea
\mfH:=\left[1+\pa_v(v\bar v\mfF)\right]^2+\bar v q\pa_q\mfF\left[1+\pa_v(v\bar v\mfF)\right]-2v\bar v^2(\pa_q\mfF)^2~.
\eea
It should be noted that in deriving (\ref{W^2}) we have made use of the identity
\bea
\h^2(\cD_\a\h^\b)(\cD^\a\h_\b)=4v\h^2~.
\eea
Eq. (\ref{W^2}) and (\ref{EoM}) imply that
\bea
u &\approx & v\mfH~,\label{ueffv}\\
p &\approx & q\left[1+\pa_v(v\bar v\mfF)\right]-4v\bar v\pa_q\mfF\label{peffq}
\eea
respectively. The symbol $\approx$ is used to indicate that the result holds modulo terms proportional to $\h_\a$ and $\bar\h_\ad$ (or, equivalently, to $W_\a$ and $\bar W_\ad$). The equations (\ref{ueffv}) and (\ref{peffq}) are the ``effective'' relations used to relate the auxiliary variables $(v,\bar v,q,\bar q)$ to the original multiplet variables $(u,\bar u,p,\bar p)$. 

The identities (\ref{auxidentities}) may be used to derive the following integral relations
\begin{subequations}\label{integralrelations}
	\bea
	\int{\rm d}^4 x \rd^2\q\,\cE\, \h W &=& \int{\rm d}^4 x \rd^2\q\,\cE\, \h^2 \non\\
	&&-\frac{1}{2}\int \rd^4 x \rd^2 \q \rd^2\bar \q \,E \, \eta^2 \bar \eta^2\Big(\pa_v(v\mfF)+\frac{1}{2}q\pa_q\mfF\Big)~,\\
	\int{\rm d}^4 x \rd^2\q\,\cE\, W^2 &=& \int{\rm d}^4 x \rd^2\q\,\cE\, \h^2 \non\\
	&&-\int \rd^4 x \rd^2 \q \rd^2\bar \q \,E \, \eta^2 \bar \eta^2\Big\{\pa_v(v\mfF)+\frac{1}{2}\bar v[\pa_v(v\mfF)]^2 \non\\
	&&+\frac{1}{2}q\pa_q\mfF(1+\pa_v(v\bar v\mfF))-v\bar v(\pa_q\mfF)^2\Big\}~.
	\eea
\end{subequations}
These relations along with (\ref{W^2Wbar^2}) allow us to rewrite the action (\ref{auxiliaryaction}) in terms of the vector multiplet,
\bea
S[W,\bar W]&=&\frac{1}{4}\int{\rm d}^4 x \rd^2\q\,\cE\,W^2 +{\rm c.c.}+\frac{1}{4} \int \rd^4 x \rd^2 \q \rd^2\bar \q \,E \, W^2 \bar W^2 \L(u,\bar u,p,\bar p)~,
\eea
where we have introduced
\bea\label{self-intmult}
\L(u,\bar u,p,\bar p):=\frac{\mfF+\mfG+\bar\mfG}{\mfH\bar\mfH}~,
\eea
and
\bea
\mfG:=\bar v[\pa_v(v\mfF)]^2+q\pa_q\mfF\pa_v(v\bar v\mfF)-2v\bar v(\pa_q\mfF)^2~.
\eea

The super-Weyl invariant version of the model (\ref{auxiliaryform}) is given by
\bea\label{superweylauxiliaryaction}
S[W,{\bar W},\h,\bar\h;\U] &=&
\int{\rm d}^4 x \rd^2\q\,\cE \Big\{ \eta W -\hf \eta^2 - \frac{1}{4} W^2\Big\} 
+{\rm c.c.}\non\\
&& + \frac{1}{4}  \int \rd^4 x \rd^2 \q \rd^2\bar \q \,E \,
\frac{\h^2\,\bar \h^2}{\U^2}\mfF\left(\frac{v}{\U^2},\frac{\bar v}{\U^2},\frac{q}{\U},\frac{\bar q}{\U}\right)~,
\label{auxiliarySG}
\eea
in which the auxiliary variable $\h_\a$ transforms as
\bea
\d_\s\h_\a=\frac{3}{2}\s\h_\a~,
\eea
in conjunction with the transformation of $\U$, eq. (\ref{compensatortrans}).

The condition of $\sU(1)$ duality invariance (\ref{SDE1}) in the auxiliary variable formalism is equivalent to \eqref{3.4}. For the model \eqref{auxiliaryaction} this means 
manifest $\sU(1)$ invariance of the auxiliary interaction function $\mfF$ and thus
\bea\label{dualinvint}
\mfF(v,\bar v,q,\bar q)=
f \Big(\frac{v}{q^2} , \frac{\bar v}{\bar q^2} , q\bar q \Big) ~.
\eea

To demonstrate how the construction discussed above works, we provide here a simple example. Consider the following $\sU(1)$-invariant auxiliary interaction
\bea
\mfF^{(0)}(v,\bar v,q,\bar q)=\frac{\k}{v\bar v}~,\qquad \k\in\dsR~.
\eea
The effective relations (\ref{ueffv}) and (\ref{peffq}) are trivial,
\bea
u \approx  v~,\qquad p \approx  q~.
\eea 
The self-interaction defined by (\ref{self-intmult}) takes the form 
\bea
\L^{(0)}(u,\bar u,p,\bar p)=\frac{\k}{u\bar u}~,
\eea
which coincides with the $n=0$ case of (\ref{self-coupling}) when $\k=\z$.
The corresponding action 
\bea
S[W,{\bar W}] &=&
\frac{1}{2} \int  \rd^4 x \rd^2 \q  \,\cE \, W^2 + \frac{\k}{4}  \int \rd^4 x \rd^2 \q \rd^2\bar \q \,E \,
\frac{W^2\,\bar W^2}{ u \bar u}
\label{3.23}
\eea
contains a cosmological term,
\bea
\k \int \rd^4 x \,e ~, \qquad e = \det (e_m{}^a) ~,
\eea
at the component level, where $e_m{}^a (x) $ is the vielbein. 

Model \eqref{3.23} is a new solution to the self-duality equation \eqref{1.3} with 
$\L(u, \bar u) = \k (u\bar u)^{-1}$, where $\k$ is the coupling constant. It was not derived in the early 2000s since Refs. \cite{KT1,KT2} studied only those models for supersymmetric nonlinear electrodynamics which possess a weak field expansion. 


\section{Superconformal duality-invariant models} \label{Section4}

As an example of applying the formalism described in the previous section,
we now turn to deriving a new superconformal duality-invariant model.
 The duality-invariant model defined by eqs. (\ref{superweylauxiliaryaction}) and (\ref{dualinvint}) is superconformal if the action is independent of $\U$. 
The most general duality-invariant and superconformal model is described by 
\bea
\mfF_{\rm SC}(v,\bar v,q,\bar q)= \frac{1}{q \bar q} \vf \Big( \frac{v}{q^2} , \frac{\bar v}{\bar q^2}  \Big) ~,
\label{4.1}
\eea
for some function $\vf ( z, \bar z)$. Choosing $\vf ( z, \bar z) = {\k}/\sqrt{z \bar z}$, 
with  $\k\in\dsR$, gives 
the model  \cite{K21}
\bea
{\mathfrak F}_{\rm SC} ( v , \bar v) = \frac{\k}{\sqrt{v \bar v}}~, 
\eea
which is the only member of the family \eqref{4.1} without dependence on $q$ and $\bar q$. 
Eliminating the auxiliary chiral $\eta_\a$ and antichiral $\bar \eta_\ad$ variables leads to the supersymmetric ModMax theory \cite{BLST2,K21}.

Here we will study a different duality-invariant and superconformal model defined by 
$\vf ( z, \bar z) = {\k}/ (z \bar z)$, with a real coupling constant $\k$, which leads to 
\bea
\mfF_{\text{SC}}(v,\bar v,q ,\bar q)=\k\frac{q\bar q}{v\bar v}~.
\eea
In this case the effective relations (\ref{ueffv}) and (\ref{peffq}) become
\bea
u &\approx & v+\k q\bar q-2\k^2{\bar q}^2~,\qquad \bar u \approx \bar v +\k q\bar q-2\k^2 q^2~,\label{ueffv1}\\
p &\approx & q-4\k\bar q~,\qquad \bar p \approx \bar q-4\k q~.\label{peffq1}
\eea 
Using the effective relations (\ref{peffq1}), we can express the auxiliary variables $q$ and $\bar q$ in terms of the multiplet variables $p$ and $\bar p$,
\bea\label{qexpp}
q \approx \frac{p+4\k \bar p}{1-(4\k)^2}~,\qquad \bar q \approx \frac{\bar p+4\k p}{1-(4\k)^2}~.
\eea
Substituting these expressions (\ref{qexpp}) into (\ref{ueffv1}) allows one to express the remaining auxiliary variables $v$ and $\bar v$ purely in terms of the multiplet variables as
\bea\label{vexpu}
v & \approx & u+\frac{4\k^2p^2(8\k^2-1)-2\k^2{\bar p}^2-\k p\bar p}{(1-(4\k)^2)^2}~,\non\\
\bar v & \approx & \bar u+\frac{4\k^2 {\bar p}^2(8\k^2-1)-2\k^2 p^2-\k p\bar p}{(1-(4\k)^2)^2}~.
\eea
With the aid of these relations (\ref{qexpp}) and (\ref{vexpu}), we can read off the self-interaction (\ref{self-intmult}) as a function of the multiplet variables,
\bea\label{self-intmod}
\L(u,\bar u,p)=\left(\frac{\k}{1-4\k}\right)\frac{p^2}{u\bar u}~.
\eea
It should be noted that for the purposes of our analysis, we have treated $p$ and $\bar p$ as independent, and only at the end of the calculation is the Bianchi identity (\ref{eq:bianchi}) imposed.

The model derived above (\ref{self-intmod}) corresponds to the $n=2$ case of the family of duality-invariant solutions in (\ref{2.19a}) with $p=\bar p$ and $\z=\k/(1-4\k)$. The outcome of our analysis is the superconformal duality-invariant model,
\bea
S[W,\bar W]=\frac{1}{4}\int{\rm d}^4 x \rd^2\q\,\cE\,W^2 +{\rm c.c.}+\frac{\z}{4} \int \rd^4 x \rd^2 \q \rd^2\bar \q \,E \, W^2 \bar W^2 \frac{(\cD W)^2}{u\bar u}~.
\eea
It is a member of the family of the superconformal vector multiplet  models \cite{K19}
\bea
S [W, \bar W] &=&
\frac{1}{4}  \int \rd^4 x \rd^2 \q   \, \cE\,
W^2 + {\rm c.c.} \non \\
&& + \frac 14\int \rd^4 x \rd^2 \q  \rd^2 \bar{\q} \, E\,
\frac{W^2\,{\bar W}^2}{(\cD W)^2}\,
\L \left( \frac{u  }{(\cD W)^{2}},
\frac{ \bar u }{(\cD W)^{2}} \right)~,
\label{action}
\eea
where $\L (\o, \bar \o)$ is a real function of one complex variable. If $\L(\o, \bar \o) $ is a solution of the self-duality equation \eqref{GZ4}, then replacing 
$
\cD W \to \hf (p+\bar p) 
$
in the action \eqref{action} leads to a superconformal duality-invariant theory.


\section{Discussion} \label{section5}

In this paper we have derived for the first time the family of solutions of the self-duality equation \eqref{SDE2.8} with $\pa_p \L \neq 0$. They have the general form 
\begin{subequations}
\bea
\L\big(u,\bar u,p,\bar p\big):={\bm \L}\big(u,\bar u; \hf p +\hf \bar p\big)
\label{5.1}
\eea 
where ${\bm \L}(u,\bar u; \g)$ is a solution of the self-duality equation \eqref{1.3} depending on a duality-invariant parameter $\g$, the latter may be a superfield such as the supergravity compensator. 
Less interesting solutions of \eqref{SDE2.8} are of the form
\bea
\widetilde{\L}\big(u,\bar u,p,\bar p\big):={\bm \L}\big(u,\bar u; \g + \frac{\ri}{2}  p -\frac{\ri}{2} \bar p\big)~.
\label{5.1b}
\eea 
\end{subequations}
Unlike \eqref{5.1}, $\widetilde{\L}\big(u,\bar u,p,\bar p\big)$ reduces to ${\bm \L}(u,\bar u; \g)$ for $p=\bar p$.
It remains an open problem to derive solutions  of the self-duality equation \eqref{SDE2.8} with 
$\pa_p \L \neq \pm \pa_{\bar p} \L$. 

In our opinion, the most interesting solutions in the family \eqref{5.1} are 
 \bea
\L\big(u,\bar u,p,\bar p\big):=\L(u,\bar u) + \frac{\x}{16} \frac{ p + \bar p}{u \bar u}~,
\eea 
where $\L(u,\bar u) $ is a solution of \eqref{1.3} and $\x$ a constant parameter. 
For $\L(u,\bar u) \neq 0$, such solutions generate new duality-invariant models for spontaneously broken local supersymmetry described by actions of the form
\bea
S = S_{\rm SG} &+& \frac 14
\int  \rd^4 x \rd^2 \q  \,\cE \, W^2 
+{\rm c.c.}
+ \frac{1}{4}  \int \rd^4 x \rd^2 \q \rd^2\bar \q \,E \,
\frac{W^2\,\bar W^2}{\U^2} \L\left(\frac{u}{\U^2},\frac{\bar u}{\U^2}
\right) \non \\
&+& \frac{\x}{8}  \int \rd^4 x \rd^2 \q \rd^2\bar \q \,E \, \U
\frac{W^2\,\bar W^2} {u\bar u} \cD^\a W_\a~,
\label{5.3}
\eea
where $S_{\rm SG}$ is the action for off-shell supergravity coupled to matter multiplets,  and $\U$ the corresponding compensator. 
Within the old minimal formulation for $\cN=1$ supergravity (see, e.g., \cite{FGKV}), $\U$ is given by 
\bea
\U = \bar S_0  \re^{-\frac{1}{3} K(\f , \bar \f)}S_0 ~, 
\label{5.4}
\eea
where $S_0$ is the chiral compensator, and $K(\f ,\bar \f)$ 
 is the K\"ahler potential for a  K\"ahler-Hodge manifold in which the matter chiral superfields $\f$ take their values.  
The matter-coupled supergravity action is
\bea
S_{\rm SG} &=& - 3
\int {\rm d}^{4} x \rd^2\q\rd^2\bar\q\,
E\,  \bar S_0  \,
\re^{-\frac{1}{3} K(\f , \bar \f)}S_0
+ \Big\{     \int {\rm d}^{4} x \rd^2 \q\,
{\cal E} \, S_0^3   W(\f)
+ {\rm c.c.} \Big\}~.
\eea
The fact that $\U$ should have the form \eqref{5.4} to preserve the K\"ahler invariance, was first pointed out in \cite{ACIK2,AKK}, see also \cite{K19}.

The supergravity-matter system \eqref{5.3} with $\L(u,\bar u)=0$  was proposed in \cite{Cribiori:2017laj} (in conjunction with the clarifying comments given in  \cite{ACIK2,AKK}).
In the $\L(u,\bar u)\neq 0$ case, this $\sU(1)$ duality-invariant theory is new, to the best of our knowledge.

In the main body of this paper, we concentrated on generating solutions to the self-duality equation \eqref{SDE2.8} without worrying about consistency of such duality-invariant models on the mass shell. The main technical 
issue here is related to the coupling \eqref{self-coupling2}.
In order for such a coupling to be well-defined, the descendant $u$ is required to be nowhere vanishing, and this requirement should be consistent with the equations of motion.
To discuss this issue, it suffices to consider a flat superspace background. We introduce the component fields of the vector multiplet following \cite{KT2}
\bea
W_{\a}\arrowvert = \j_{\a}~,\qquad
-\frac{1}{2} D^{\a}W_{\a}\arrowvert=D~,
\qquad
D_{(\a}W_{\b)}\arrowvert 
=2 {\rm i} {F}_{\a\b}
&=& {\rm i} (\s^{bc})_{\a\b}{F}_{bc}~,
\eea
where the bar-projection  $U|$ of a superfield $U$ means, as usual, switching off the superspace Grassmann variables, see e.g. \cite{Ideas,GGRS}. It holds that 
\bea
u| \equiv \frac 18 D^2 W^2|  ={\bm u}+
\text{fermionic terms}~, \qquad {\bm u} =F^{\a\b} F_{\a\b} -\hf D^2 = \o - \hf D^2~,
\eea
where $\o$ is given by \eqref{omega}.
For the coupling \eqref{self-coupling2}, the vector multiplet  action in the Minkowski background is
\bea
S[W,{\bar W} ] &=&
\frac{1}{4} \int  \rd^4 x \rd^2 \q  \, W^2 +{\rm c.c.} + \frac{1}{4}  \int \rd^4 x \rd^2 \q \rd^2\bar \q \,
\frac{W^2\,\bar W^2} {u\bar u} {\mathfrak D}\big( D^\a W_\a \big) \non \\
&=&  \int \rd^4 x \,\left\{
-\frac 14 F^{ab}F_{ab} + \hf D^2 + {\mathfrak D}\big( -2D \big) \right\} + \text{fermionic terms}~.
\eea
All the non-analytic contributions are concentrated in the fermionic sector. 
Since the equation of motion for $D$ is 
\bea
D = 2 {\mathfrak D}'\big( -2D \big) + \text{fermionic terms}~,
\label{5.9}
\eea
there are two ways for $\bm u$ to be nowhere vanishing. The first option is that  \eqref{5.9} has a non-zero solution $D_0 \neq 0$, which is the case for 
${\mathfrak D}(y)= \frac{\z}{8} y^2 +\frac{\x}{2} y$. 
The second option is realised when the solution to \eqref{5.9} is 
$D=0$ modulo fermionic contributions, and then $F^{\a\b} F_{\a\b} $ should be restricted to be nowhere vanishing, as  in the ModMax theory \cite{BLST}.

The bosonic sector of the supergravity-matter system \eqref{5.3} with $\x=0$ (coupled to the dilaton-axion multiplet) was computed in \cite{KMcC2}, to which the reader is referred for the details. Switching off the supergravity multiplet in \eqref{5.3}, including the compensator, the bosonic action takes the form
\bea
S_{\rm boson} =  \int \rd^4 x \, \cL~, \qquad 
\cL=-\frac 14 F^{ab}F_{ab} + \hf D^2 + {\bm u} \bar {\bm u} \L \big({\bm u}, \bar {\bm u} \big)  - \x D~,
\label{5.10}
\eea
and the equation of motion for $D$ becomes \cite{K-FI}
\bea
D \Big[1- \bar {\bm u} \G ({\bm u} , \bar {\bm u}) -  {\bm u}\bar \G  ({\bm u} , \bar {\bm u}) \Big]
= \x~.
\label{D2}
\eea 
Generically, the auxiliary field develops a non-vanishing expectation value, 
$\langle D \rangle \neq 0$, which must satisfy an algebraic nonlinear equation that 
follows from (\ref{D2}) by setting $F^{\a\b} F_{\a\b} =0$. 
As a result, the supersymmetry becomes spontaneously broken.
For example, the $\cN=1$ supersymmetric Born-Infeld action  \cite{CF,BG,RT} 
 is described by 
\bea
\L_{\rm SBI}(u,\bar u; g) = \frac{g^2}{ 1 + \frac 12 {g^2} (u+\bar u)  +
\sqrt{1 + g^2 (u+\bar u)  +\frac 14 g^4 (u-\bar u)^2} }~.
\eea
The corresponding bosonic Lagrangian density, eq.  \eqref{5.10}, can be written in the form
\bea
\cL_{\rm SBI}=   \frac{1}{g^2} \bigg\{ 
1 - \sqrt{1 + g^2 ({\bm u} + {\bar {\bm u}} )
+{1 \over 4}g^4 ({\bm u} - {\bar {\bm u}} )^2 } 
\bigg\} - \x \, D ~.
\eea 
In this case the equation \eqref{D2} is solved as \cite{K-FI}
\bea
D = \frac{\x}{
\sqrt{1+ g^2 \x^2}
} \,
\sqrt{1 + g^2 (\o + {\bar \o} )+{1 \over 4}g^4 (\o - {\bar \o} )^2 } ~.
\label{D3}
\eea

In general, the component reduction of the supergravity-matter system \eqref{5.3} may be obtained by applying the technique developed in \cite{BLST2} using the earlier construction of \cite{CFT}. This will be discussed elsewhere.
\\

\noindent
{\bf Acknowledgements:}\\
We thank the referee of this paper for constructive suggestions.  
We are grateful to Emmanouil Raptakis for useful comments on the manuscript.
The work of SMK is supported in part by the Australian 
Research Council, projects DP200101944 and DP230101629.
The work of JCS is supported by the Australian Government Research Training Program Scholarship.


\begin{footnotesize}

\end{footnotesize}


\end{document}